\documentclass[prl,twocolumn,showpacs]{revtex4-1}
\usepackage[xdvi]{graphicx}%
\usepackage{dcolumn}
\usepackage{amsmath}

\newcommand{\bra}{\langle}
\newcommand{\ket}{\rangle}

\newcommand{\bv}[1]{{\boldsymbol #1}}

\begin{document}

\title{Pure Glass in Finite Dimensions}

\author{Shin-ichi Sasa}
\affiliation{Department of Basic Science,
The University of Tokyo, Tokyo, 153-8902, Japan}
\date{\today}

\begin{abstract}
Pure glass is defined as a thermodynamic phase 
in which typical equilibrium particle configurations 
have macroscopic overlaps with one of some special
irregular configurations. 
By employing 128-types of artificial molecules, 
a pure glass model is constructed in the cubic 
lattice. 
\end{abstract}



\maketitle


\paragraph{Introduction}


Structural glass transitions have been extensively studied 
from several viewpoints. Among various approaches, 
the random first-order transition scenario (RFOT) 
can be a candidate theory for the mean-field description 
of supercooled liquids and glasses 
\cite{KTW,replica,ParisiZamponi,RFOT}. However, the 
precise nature of  glass in finite dimensions has not
been clarified.  For example, it is a long standing question
whether a thermodynamic transition to a pure glass phase 
exists or not \cite{Kauzmann}.
Why is such an apparently simple question difficult? 
In laboratory and numerical experiments, true equilibrium 
properties in  prospective glassy materials are hardly
observed, because the equilibration time for a glass
phase is considerably longer than accessible time  scales. 
Theoretically,
a renormalization group analysis may be a suitable approach 
to solving the above simple question \cite{RG,Yeo}, 
but the agreement between the two studies has not been obtained.
In this Letter, an attempt is made to answer  the question of
the existence of pure glass in finite dimensions by adapting 
a different approach. For this purpose, a slightly 
complicated model is constructed first, and then 
theoretical arguments concerning the model are presented. 


The first step in this approach is to assume statistical 
mechanical conditions for constructing a pure glass. 
It is believed that particles in the glass phase are frozen 
into  some irregular configurations. This hypothesis is 
expressed by the following condition: typical equilibrium 
particle configurations have {\it macroscopic overlaps with 
one of  some special irregular configurations} (MOSIC). 
The condition 
provides a precise definition of  pure glass in this Letter. 
It should be noted that pure glass presented thus far are limited 
to models in a random graph \cite{BiroliMezard,Coniglio,Krzakala}
or with long-range interactions in the 
Kac limit \cite{Franz,FranzMontanari}. 
One of the main results  
in this  Letter is the presentation of a pure glass model
with short-range interactions in finite dimensions.


At this point, one may ask whether  pure glass can be 
regarded as an idealization of glassy materials in  nature. 
Furthermore, one may be interested in knowing whether 
the mode coupling theory \cite{MCT} and the RFOT 
provide a useful description of the dynamics and thermodynamics 
of pure glass. An immediate answer to these questions is 
not available. Nevertheless, since pure glass is different
from  gas, liquid, crystal, and  quasi-crystal,
the study on pure glass might shed  light on  the 
understanding of glassy materials and stimulate theoretical
studies on glasses from a different perspective.



\paragraph{Model}

\begin{figure}[htbp]
\begin{center}
\includegraphics[width=8cm]{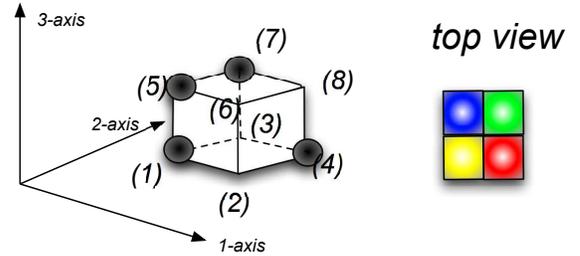}
\end{center}
\caption{Unit cube with marked vertices. In the figure
on the left-hand  side, $\sigma^{(k)}=1$ only for $k=1,4,5,7$.
The figure on the right-hand side represents the top view of 
the cube, where one of the four colors is assigned for each pair 
of ($\sigma^{(k)}$,$\sigma^{(k')}$) aligned in the vertical direction.  
For example, red is assigned 
for $(\sigma^{(2)}=0$,$\sigma^{(6)}=0)$.}
\label{cube}
\end{figure}

If an infinite series of local minimum configurations (LMCs) 
in a model are understood theoretically, the statistical behavior of the 
model may be conjectured on the basis of an energy landscape  of LMCs. 
The proposed model, where 128-types of artificial molecules
are considered, is  constructed along with this basic concept. 
Each molecule type is represented by an integer $\sigma$
in $\{0,1,\dots,127 \}$. Given $\sigma$, the binary expansion 
$ \sigma = \sum_{k=1}^7 \sigma^{(k)}2^{k-1} $
uniquely defines  $\sigma^{(k)}\in\{0,1\}$, $1 \le k\le 7$.
Further, a ``sufficiently irregular'' binary-array 
$ (\sigma^{(8)}(\sigma))_{\sigma=0}^{127}$ is picked and fixed.
(Such an array was generated by selecting the value of 
$\sigma^{(8)}(\sigma)$ as $0$ or $1$ with probability $1/2$ 
for each $\sigma$.) As shown in Fig. \ref{cube}, 
the molecule type $\sigma$ corresponds to a unit 
cube whose $k$-th vertex is marked only when 
$\sigma^{(k)}=1$.  
Molecules are interpreted as  three-dimensional
generalization of Wang tiles \cite{wtile,Robinson},
where a  mark configuration  such as 
$(\sigma^{(1)}, \sigma^{(2)},\sigma^{(5)},\sigma^{(6)})$
on each plaquette represents the  `` color''  of the plaquette.


\begin{figure}[htbp]
\begin{center}
\includegraphics[width=8cm]{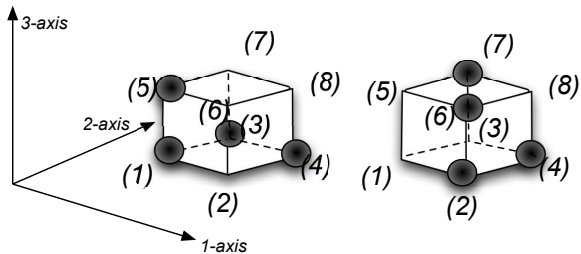}
\end{center}
\caption{Example of the interaction energy.
For two cubes $\sigma=29$ (left) and $\sigma'=106$ (right),
$V_1(\sigma,\sigma')=1$, $V_2(\sigma,\sigma')=1$, 
$V_3(\sigma,\sigma')=1$, $V_2(\sigma',\sigma)=1$,
and $V_3(\sigma',\sigma)=1$. $V_1(\sigma',\sigma)$
is either $0$ or $-1$, which is selected with 
probability $1/2$.}
\label{int}
\end{figure}

It is assumed that only one molecule can occupy 
each site in  the cubic  
lattice $\Lambda=\{i \in {\bf N}^3|
i=(i_1,i_2,i_3),  1 \le i_k \le  L \}$. 
The total number of sites in the lattice is $N=L^3$.
A molecule configuration is denoted by 
$\bv{\sigma}=(\sigma_i)_{i \in \Lambda}$. 
For any nearest neighbor pair of sites $(i,j)$
that satisfies the relation $j-i= e_k$, where $e_k$ 
is the unit vector in the $k$ direction of the lattice,
the interaction energy $V_{k}(\sigma,\sigma')$ is
defined for $(\sigma_i,\sigma_j)=(\sigma,\sigma')$
as follows: $V_{k}(\sigma,\sigma')=1$ when mark configurations 
in the adjacent plaquettes of unit cubes at sites 
$i$ and $j=i+e_k$ are different; otherwise, 
$V_{k}(\sigma,\sigma')$  takes either $0$ or $-1$
irregularly. Such an irregular function was generated
by selecting the value $0$ or $-1$ with probability $1/2$
for each case. The irregularity of $V$ is necessary to 
introduce a rugged energy landscape of LMCs.
An example of the 
interaction energy is shown in Fig. \ref{int}.
The Hamiltonian of the proposed model is given by
\begin{equation}
H(\bv{\sigma} ) 
= \sum_{\bra i,j\ket} V_{k}(\sigma_i,\sigma_j),
\end{equation}
where $\bra i,j \ket$ represents a  nearest neighbor
pair of sites. Although some elements of the interaction 
energy are determined by using probabilities, a 
quenched disorder does not exist in 
the lattice. Indeed, $H(\bv{\sigma} ) $ is translational
invariant if the periodic boundary condition is assumed.
Throughout this Letter, $\beta$ represents  inverse temperature.


\paragraph{Analysis}


A molecule configuration without any mismatches between 
the mark configurations in all adjacent 
plaquettes is called a  perfect matching configuration (PMC).
The advantage of the proposed model is  that all PMCs can be constructed 
using a simple iteration rule. The rule is as follows.
First, the values of  $\sigma_i^{(k)}$ for  $k=1,2,3,4$ 
in plane  $i_3=1$; for  $k=1,2,5,6$ in plane $i_2=1$; 
and for  $k=1,3,5,7$ in plane $i_1=1$ 
are randomly chosen. Then, the molecule type $\sigma_i$ 
for $i=(1,1,1)$ is determined. Accordingly, the value 
of $\sigma_{i}^{(8)}$ for $i=(1,1,1)$ is given and 
this value is equal to the value of $\sigma_{i}^{(7)}$ 
for $i=(2,1,1)$.
Thus, the molecule type $\sigma_i$ for $i=(2,1,1)$ 
can be determined.  By repeating  this procedure, 
all the values of $\sigma_i$ for $i \in \Lambda$ 
can be determined  depending on the initial choice of 
mark configurations in the planes. 
Since all PMCs are generated using the iteration rule,
there are $2^{3 L^2 +3 L+1}$ PMCs.

\begin{figure}[htbp]
\begin{center}
\includegraphics[width=4cm]{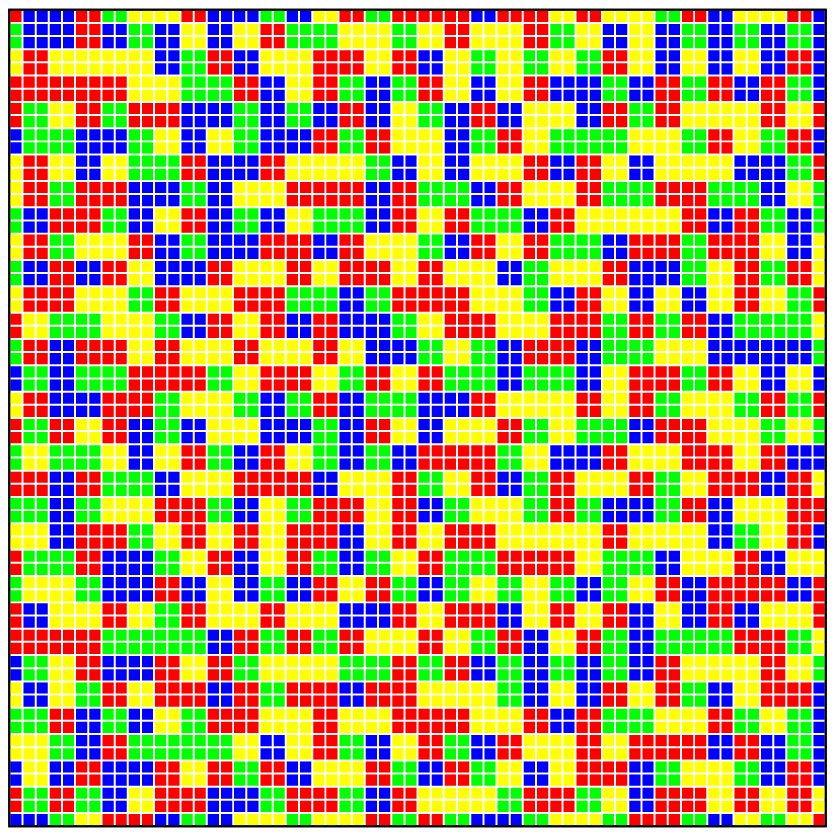}
\end{center}
\caption{Perfect matching configuration $(L=32)$. 
Molecule configuration in  plane $i_3=L/2$ is 
shown as the color representation of the top view 
of cubes. (See Fig. \ref{cube} for the color representation.)}
\label{config}
\end{figure}

An important property of PMCs is that they exhibit
neither long range positional order nor internal symmetry 
breaking. This property is evident from Fig. \ref{config}, where 
the cube configuration in plane $i_3=L/2$ is shown for a 
randomly chosen PMC. This configuration is certainly irregular. 
One can also examine this irregularity by measuring some statistical 
quantities. Theoretically, the irregularity may be 
understood from the deterministic iteration rule used for 
constructing PMCs. Although its rigorous proof is not obtained
yet, one can demonstrate that the rule maintains the random nature
of configurations in the planes $i_1=1$, $i_2=1$, and $i_3=1$ 
(See Ref. \cite{Sasa} for a related argument.). 


Next, the energy landscape of $H(\bv{\sigma})$ is discussed.
It can be easily confirmed that the replacement of any cube
in a  PMC always increases the energy. 
That is, all PMCs are LMCs. Moreover, LMCs other than PMCs  
exist. 
Suppose that a cube at site $i$ in a PMC 
is altered such that either $\sigma_i^{(6)}$ or $\sigma_i^{(7)}$
is changed. Then, three mismatches of $\sigma_i$ 
are generated in the configuration. 
Two of them can be eliminated by employing 
the iteration rule beginning from the cube next to $\sigma_i$.
This leads to an  LMC with one mismatch 
which cannot be eliminated by  any finite number of 
steps of cube replacement in the thermodynamic limit. 
The repetition of this procedure can yield LMCs with $k$ 
isolated mismatches which cannot be eliminated by any 
finite number of steps of cube replacement in the thermodynamic 
limit, where $k$ is an arbitrary finite integer. 
A collection of all such LMCs is denoted 
by $(\bv{\sigma}^\alpha)_{\alpha \in  {\cal A}}$,
where ${\cal A}$ is the index set. $|{\cal A}|$ is not estimated, 
but from the construction method of LMCs, it is found 
that $|{\cal A}| \simeq \exp(s_0 N)$, where $s_0$ is 
a positive constant. 

Since the energy density $u_\alpha \equiv H(\bv{\sigma}^\alpha)/N$ 
for $\alpha \in {\cal A}$ is obtained as the space average of 
interaction energies for an irregular configuration $\bv{\sigma}^\alpha$, 
the distribution of $u_\alpha$ has a sharp peak with width of 
$O(1/\sqrt{N})$. Thus, as studied in the random energy model 
\cite{REM}, it is expected 
for sufficiently low temperature that the statistical measure 
condensates onto  configurations 
near a $\bv{\sigma}^\alpha $ satisfying 
$H(\bv{\sigma}^\alpha) \simeq u_* N$,
where $u_*$ is the minimum energy density of LMCs in the thermodynamic 
limit. This condensation phenomenon leads to MOSIC. ''


\paragraph{Numerical experiments}


The condensation phenomenon is explored
by numerical experiments.
The statistical quantities for rather small-size 
systems in the equilibrium state under free 
boundary conditions are calculated 
by employing the exchange Monte Carlo method \cite{HukushimaNemoto}.

\begin{figure}[htbp]
\begin{center}
\includegraphics[width=4cm]{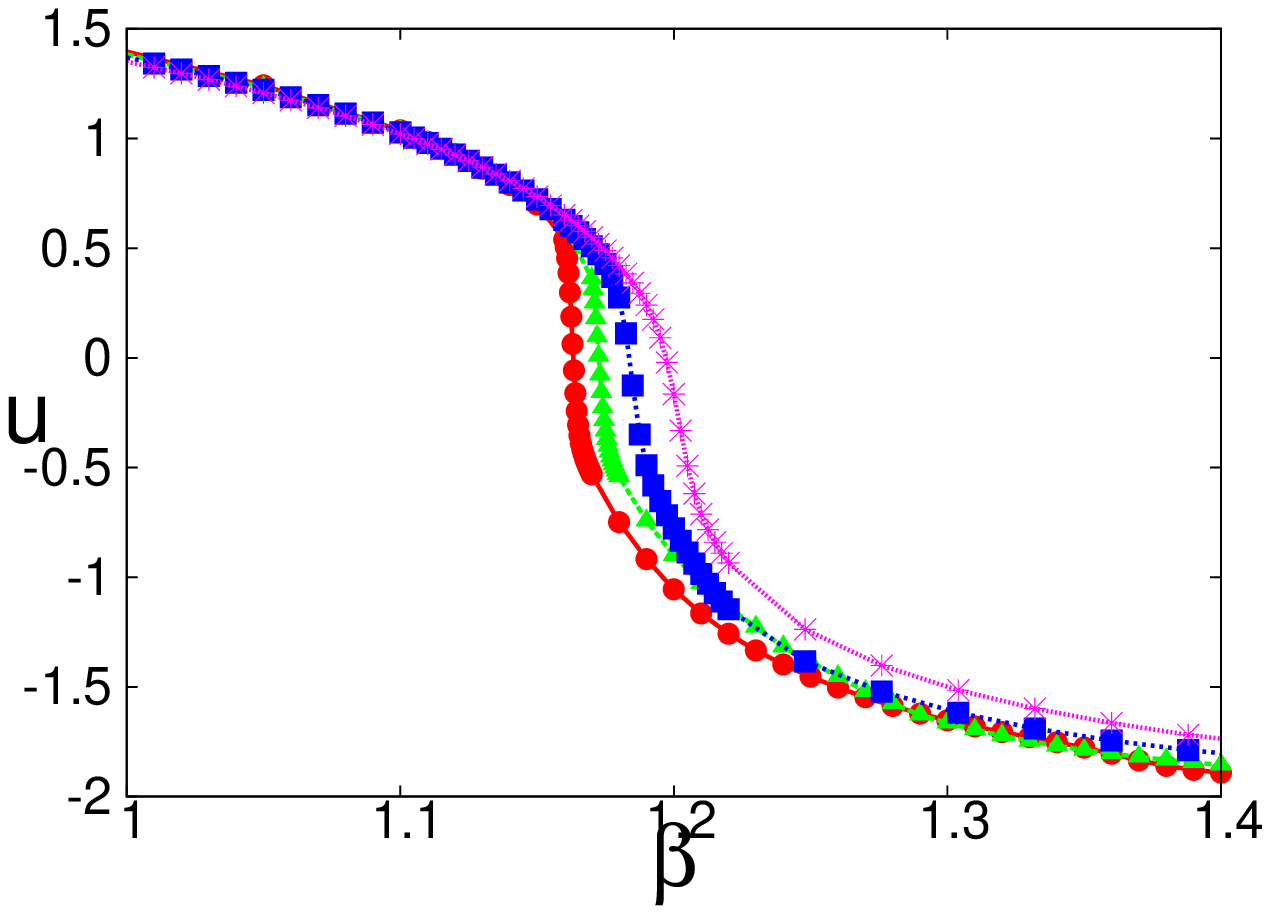}
\includegraphics[width=4cm]{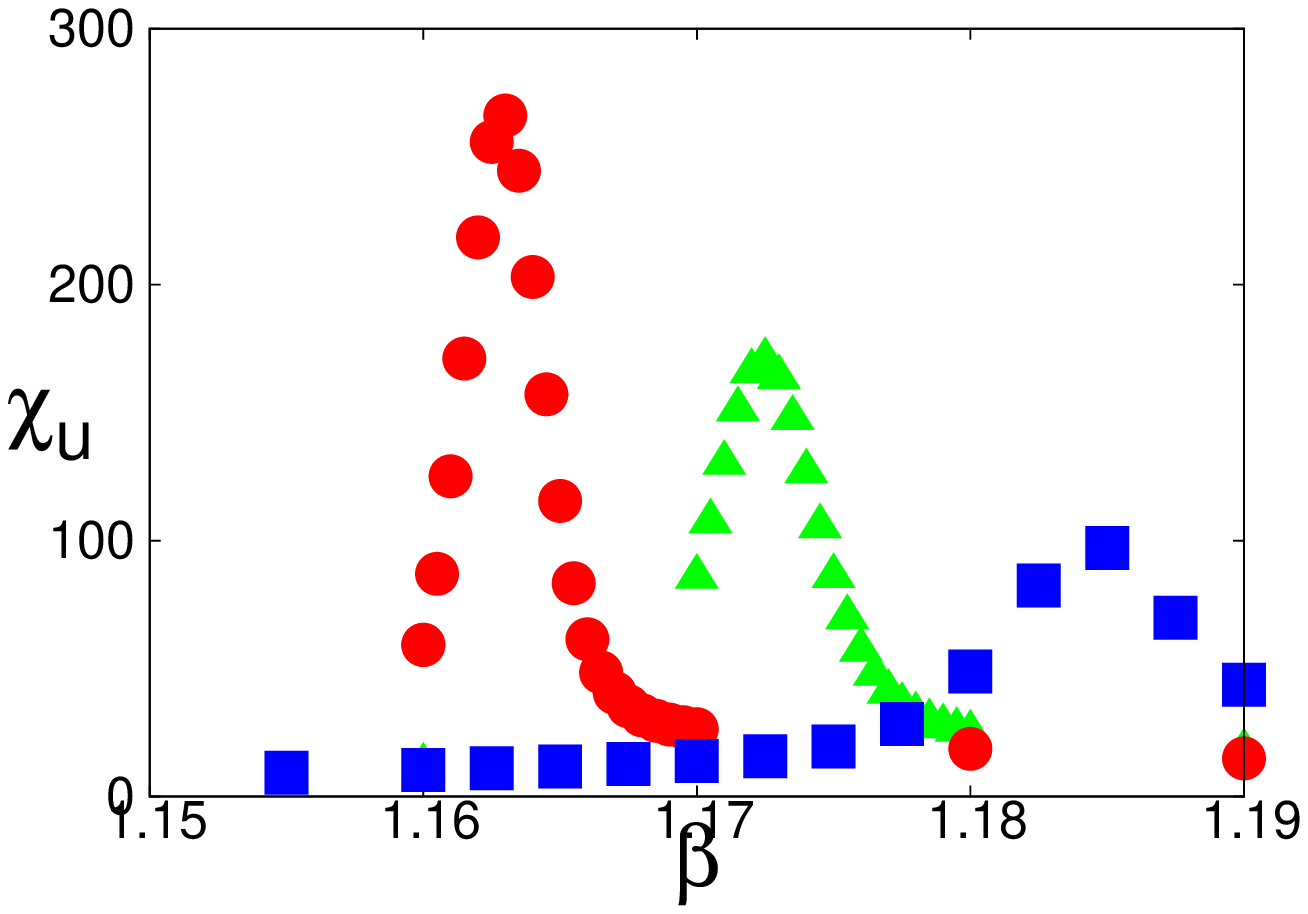}
\end{center}
\caption{(Color online) Thermodynamic quantities for several
values of $L$. The left figure shows  the energy 
density $u$ as functions of $\beta$ for 
$L=8$ (asterisk), 9 (square), 10 (triangle) and 
$11$ (circle).  The intensity of energy fluctuations
is  displayed in the right side. $L=9$ (square), 
10 (triangle) and $11$ (circle).}
\label{u-data}
\end{figure}
\begin{figure}[htbp]
\begin{center}
\includegraphics[width=4cm]{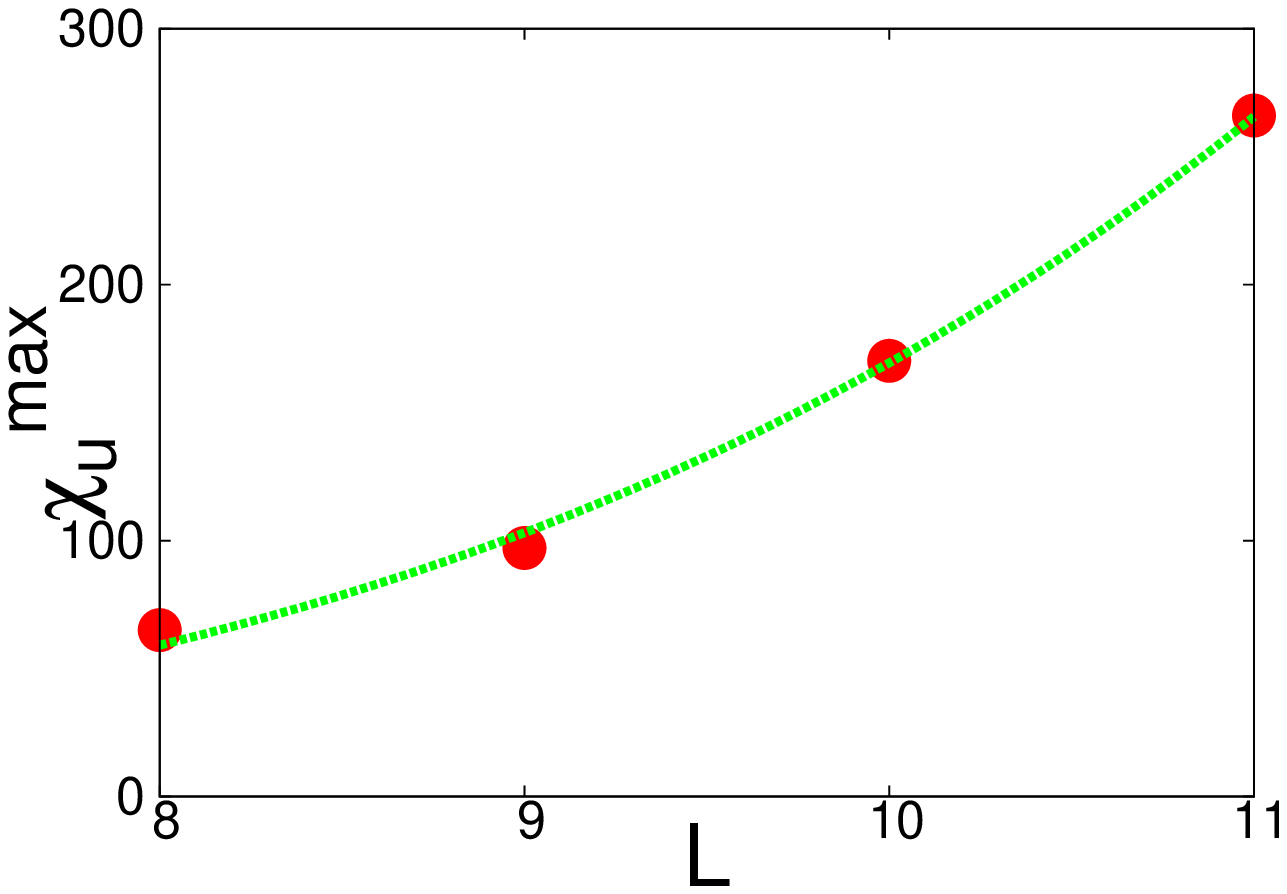}
\includegraphics[width=4cm]{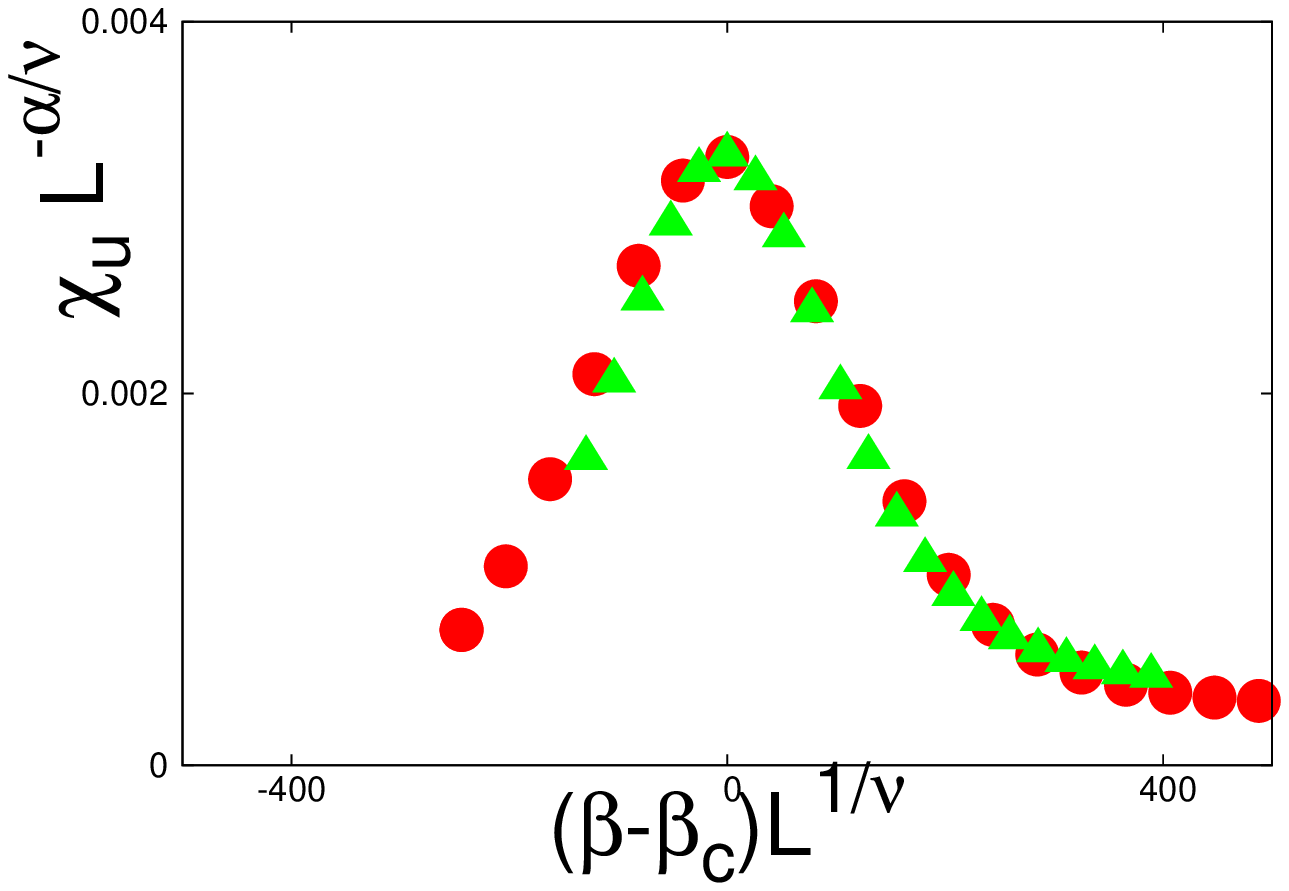}
\end{center}
\caption{(Color online) 
The left graph indicates 
the maximum value of $\chi_u$ for $L$.
The guide line represents $\chi_u = (L/3.4)^{4.7}$. 
The right figure  shows  $\chi_u L^{-4.7}$ versus 
$(\beta-\beta_{\rm c}(L))L^{4.7}$ for two graphs 
with $L=10$ and $L=11$.}
\label{peak}
\end{figure}

First, thermodynamic quantities for the proposed model are
investigated.  As an  example, in the left side of 
Fig. \ref{u-data},
the change in  the energy density $u=\bra \hat u \ket$ 
with inverse temperature $\beta$ is shown for $L=8,9,10,11$, 
where $\hat u=H(\bv{\sigma})/N$. An abrupt drop 
is observed around a  temperature for each $L$. 
Since  $-du/d \beta $ is equal to the intensity of 
energy fluctuations $\chi_u= 
N \bra (\hat u-u )^2 \ket $,  $\chi_u$ is displayed
in the right side of  Fig. \ref{u-data}. 
The peak value of $\chi_u$, denoted by 
$\chi_u^{\rm max}$, increases  as 
$\chi_u^{\rm max}\simeq L^{4.7}$ (see Fig. \ref{peak}),
which is faster than 
$\chi_u^{\rm max}\simeq L^{3}$
expected for the first-order transition. 
Furthermore, two graphs of $\chi_u$ for $L=11$ and $L=10$ 
in Fig. \ref{u-data} satisfy a relation
\begin{equation}
\chi_u \simeq L^{\alpha /\nu} f( (\beta-\beta_{\rm c}(L))L^{1/\nu}),
\end{equation} 
where $\alpha=1.0$ and $\nu=1/4.7 \simeq 0.21$.
$\beta_{\rm c}(L)$ is estimated as the value of $\beta$
where $\chi_u$ becomes maximum. 
From these results, it is certain that 
a thermodynamic transition  occurs  at a non-zero 
temperature in the thermodynamic limit,
and it may be conjectured that the transition is 
the first order. 
It should be noted that the system sizes in the experiment are 
too  small to identify the nature  of the transition precisely. 


\begin{figure}[htbp]
\begin{center}
\includegraphics[width=7cm]{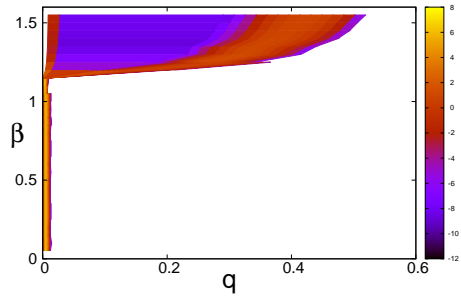}
\end{center}
\caption{A color representation of $\log P$ 
as a function of $q$ for different values of $\beta$. 
$L=10$. 
}
\label{pofq}
\end{figure}

Next, in order to observe MOSIC in low temperatures, 
two independent identical systems are prepared, 
where two  configurations are denoted by 
$(\sigma^{(1)}_i)_{i \in \Lambda}$ 
and $(\sigma^{(2)}_i)_{i \in \Lambda}$, respectively. 
The overlap between the two configurations is defined as
\begin{equation}
\hat q\equiv\frac{1}{N} 
\sum_{i \in \Lambda}\delta(\sigma_i^{(1)},\sigma_i^{(2)}).
\end{equation}
Let $P(q,\beta)$ be  the distribution function 
of $\hat q= q$ in the equilibrium state of the system with inverse 
temperature $\beta$. In Fig. \ref{pofq}, the color 
representation of $\log P(q,\beta)$ is shown, 
which  clearly indicates the existence of two peaks 
at low temperatures. Since typical configurations 
in low-temperatures are irregular, it is concluded
that MOSIC emerges. This is the most remarkable 
result in this Letter. 


\begin{figure}[htbp]
\begin{center}
\includegraphics[width=4cm]{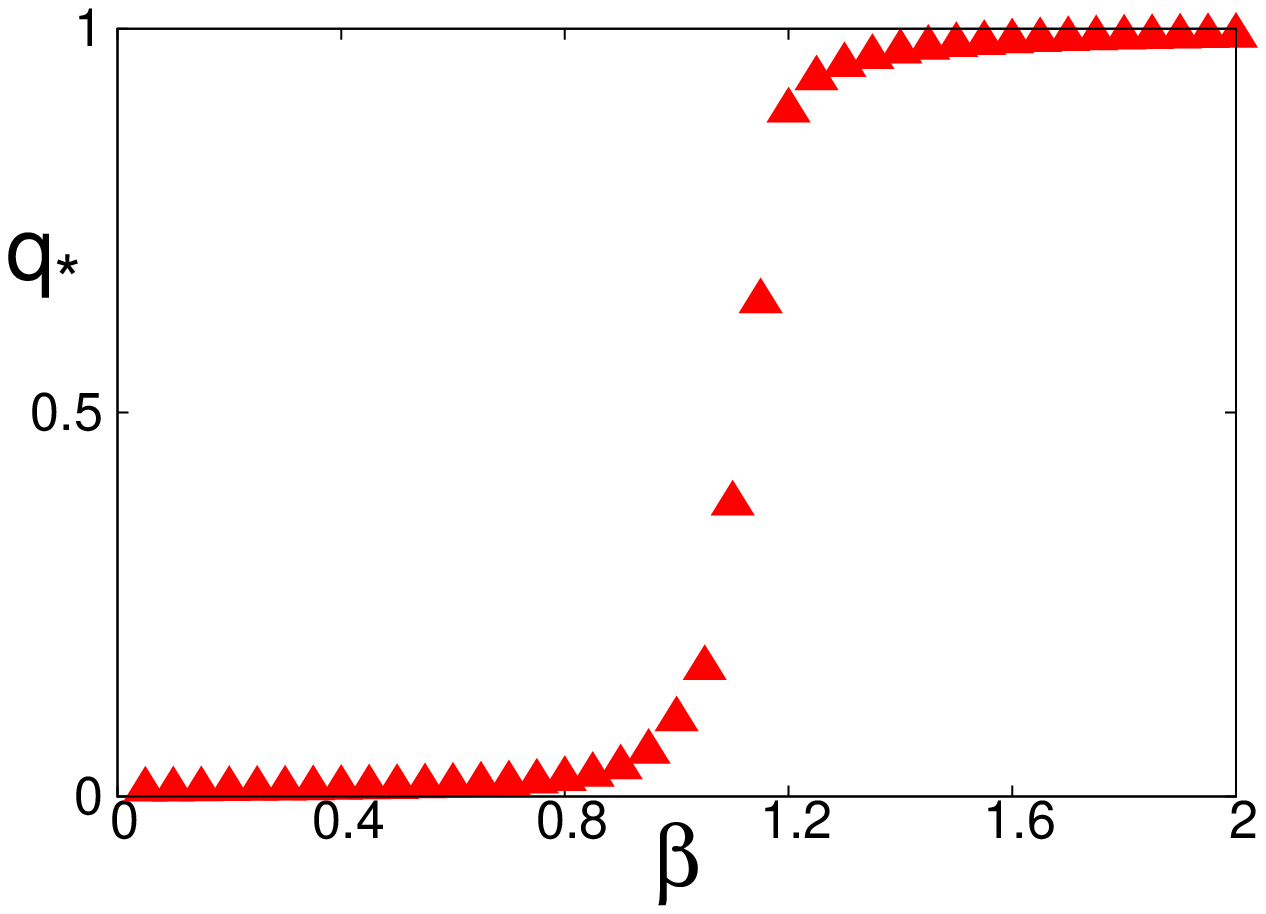}
\includegraphics[width=4cm]{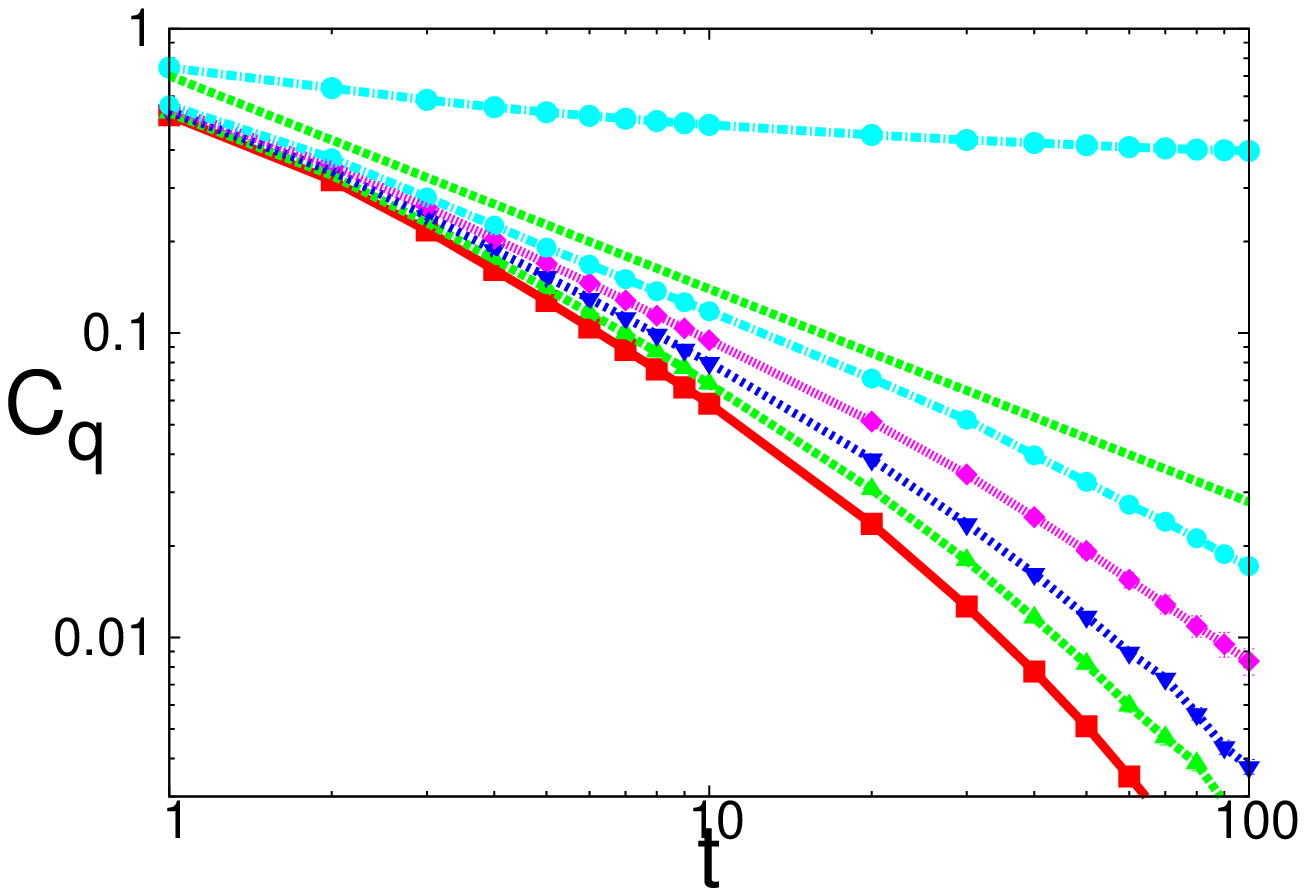}
\end{center}
\caption{(Color online) 
The left-side figure shows 
$q_* $  as a function of $\beta$ $(L=10)$. 
In the right side, $C_q(t)$ are plotted 
with a log-log scale.
$\beta= 1.15$ (squares), 
$1.155$ (upward triangles), $1.16 $ (downward triangles),  
$1.165 $ (diamonds), and  $1.17 $ (circles).
Since trajectories for $\beta=1.17$ are clearly classified
into two  groups, two graphs corresponding to the 
two groups are plotted. 
The dotted straight line represents a power-law form 
$C_q(t) = 0.7 t^{-0.7}$ as a guide line.
}
\label{add}
\end{figure}

A further evidence of MOSIC 
is proposed by studying the system under 
a special boundary configuration, 
where molecules in the boundary planes $i_k=1$ 
and $i_k=L$ ($k=1,2,3)$ are fixed as those of 
$\bv{\sigma}^*$ chosen from equilibrium configurations 
with large $\beta$. Then, as shown in the left side of 
Fig. \ref{add}, the overlap with $\bv{\sigma}^*$, 
which is characterized by 
\begin{equation}
q_*=\frac{1}{\tilde N}\sum_{i \in \tilde \Lambda} 
\bra \delta(\sigma_i,\sigma_i^*) \ket,
\end{equation}
approaches  one as $\beta$ is increased,
where $\tilde \Lambda $ represents the inner region
except for the boundaries in $\Lambda$.
$\tilde N =(L-2)^3$ is the number of sites in $\tilde \Lambda$.
The result indicates 
that the boundary configuration selects one irregular 
configuration, just as the spin-up boundary 
configuration determines the ordered phase 
with positive magnetization. This is a direct evidence 
that there exists a low-temperature phase with MOSIC.


Finally, properties near the transition point are briefly studied
from a viewpoint of dynamics. The simplest quantity characterizing
dynamical behaviors is the correlation function
\begin{equation}
C_{q}(t)=\frac{1}{N}
\sum_{i \in \Lambda} \bra \delta(\sigma_i(t),\sigma_i(0)) \ket-\frac{1}{128},
\label{cqdef}
\end{equation}
where $\sigma_i(0)$ is sampled in the  equilibrium state
and the time evolution $\sigma(t)$ obeys the Glauber dynamics
with the heat bath method. It is observed that $C_q(t)$ decreases
exponentially in sufficiently high temperature, while it 
saturates to a finite value in the low temperature phase.
Near the transition point,  two types of trajectories, which 
correspond to de-correlation and freezing, coexist. 
The behavior may be consistent with the discontinuous 
transition of $q$. Here, as shown in the right side of 
Fig. \ref{add}, no intermediate plateau is observed, 
in contrast to the RFOT scenario. Although the decay of $C_{q}(t)$ 
becomes slower as $\beta$ approaches the transition point
on the high temperature side, the divergent behavior is not 
concluded in the experiment of the small size system.


\paragraph{Concluding Remarks}


In this Letter, a pure glass model in finite dimensions
has been proposed. In this model, typical
equilibrium configurations in low temperature have 
macroscopic overlaps with one of some special irregular 
configurations. Although the behavior is similar to
that observed in the models that exhibits one step
replica symmetry breaking, the nature of the transition
seems to be  different from the RFOT.
In contrast to previous numerical studies on glass transition 
in finite dimensions \cite{Rieger,Ritort,FZ}, 
the present work has explored 
the nature of the low temperature phase more than properties 
near the transition point. The two peaks in $P(q)$ in low temperatures
may be the clearest demonstration among all existing studies 
of glassy systems in finite dimensions. 
Furthermore, the construction of an infinite series of local minimum 
configurations will provide a clue for theoretical analysis 
of statistical mechanics  in sufficiently low temperatures. 


Before ending the Letter, I provide a few remarks.
The arguments presented in this Letter 
are  based on several conjectures
whose validity is  demonstrated with the aid of 
numerical experiments. A complete theory for 
describing  pure glass will be developed in 
future work. Then, it may be significant to construct
a simpler model that exhibits a behavior similar to that of
pure glass. Furthermore, it is interesting to consider 
an effective field theory for describing pure glass, 
which might be related to the replica field theory \cite{rft}.

The ultimate goal of this study is to construct
a pure glass experimentally. Toward this goal, 
the next subject is to study a Hamiltonian system for which
the rugged energy landscape may be discussed theoretically.
It will be amazing to observe the properties of 
pure glass via molecular dynamics 
simulations. In this study, complicated shapes
of molecules will be employed so as to 
produce  interaction potentials similar 
to $V_{k}(\sigma,\sigma')$.
This challenge may be related to recent 
studies on complex structures emerging out
of  designed building  blocks with anisotropy
\cite{Glotzer,Granick}.
%
%
Here, let us recall 
a history of quasi-crystals.
The mathematical study on aperiodic but regular
tiling was conducted in the 1960s \cite{wtile},
whereas the experimental construction of quasi-crystals 
was first  reported in 1984 \cite{QC-exp}.
As an extension of this Letter, a theoretical study 
on irregular but ordered tiling will be conducted. 
I hope that future researchers will attempt to consider 
the  possibility of constructing  pure glass experimentally.

The author thanks K. Hukushima, N. Mitarai, H. Ohta,  
H. Tasaki and H. Yoshino for helpful discussions. 
The present study was supported by KAKENHI Nos. 22340109 
and 23654130, and  the JSPS Core-to-Core Program 
``International research network for non-equilibrium 
dynamics of soft matter''. 


\end{document}